\documentclass[10pt, twoside, journal]{IEEEtran}

\usepackage{graphicx}
\usepackage{amsmath}

\newcommand{\dyadic}[1]{{#1}
\setbox0=\hbox{$\scriptstyle\leftrightarrow$}
   \setbox2=\hbox{$#1$}
   \dimen0=.5\wd0 \advance\dimen0 by-.5\wd2
   \advance\dimen0 by-\wd0
   \kern\dimen0
{^{\raise.2ex\hbox{$\scriptstyle\leftrightarrow$}}}}

\newcommand{\dyadictall}[1]{{#1}
\setbox0=\hbox{$\scriptstyle\leftrightarrow$}
   \setbox2=\hbox{$#1$}
   \dimen0=.5\wd0 \advance\dimen0 by-.5\wd2
   \advance\dimen0 by-\wd0
   \kern\dimen0
{^{\raise.5ex\hbox{$\scriptstyle\leftrightarrow$}}}}

\DeclareFontFamily{U}{cmbsy}{}
\DeclareFontShape{U}{cmbsy}{m}{n}{ <5> <6> <7> <8> <9> gen * cmbsy
       <10> <10.95> <12> <14.4> <17.28> <20.74> <24.88> cmbsy10}{}
\DeclareMathAlphabet{\scb}{U}{cmbsy}{m}{n}

\begin{document}

\title{Average Transition Conditions for Electromagnetic Fields at a Metascreen of Nonzero Thickness}

\author{Edward~F.~Kuester,~\IEEEmembership{Life Fellow,~IEEE} and Enbo Liu
        \thanks{Manuscript received \today.}\thanks{E. F. Kuester is with the Department of Electrical, Computer and Energy Engineering, University of Colorado, Boulder, CO 80309 USA (email: Edward.Kuester@colorado.edu). Enbo Liu is with the Department of Mechatronics, University of Electronic Science and Technology of China (UESTC), Chengdu, China.}}

\markboth{IEEE Transactions on Antennas and Propagation}{Transition Conditions for Thick Metascreen}

\maketitle

\begin{abstract}
Using a dipole interaction model, we derive generalized sheet transition conditions (GSTCs) for electromagnetic fields at the surface of a metascreen consisting of an array of arbitrarily shaped apertures in a perfectly conducting screen of nonzero thickness. The simple analytical formulas obtained are validated through comparison with full-wave numerical simulations.
\vspace{7mm}

\end{abstract}

\section{Introduction}

In \cite{ekel}, generalized sheet transition conditions (GSTCs) describing the interaction of electromagnetic waves with periodically perforated perfectly conducting (PEC) screens (which we call \emph{metascreens}) of zero thickness were derived using a dipole-interaction (Clausius-Mossotti-Lorentz-Lorenz) approximation. In many situations, the non-zero thickness of an actual conducting screen can cause significant deviation from what is predicted by this theory. Some previous work has been carried out on the modeling of thick perforated screens \cite{matsu}-\cite{garcia}, but these present reflection and transmission coefficients or impedances rather than equivalent boundary conditions, and often only numerical results rather than analytical formulas. In \cite{eom}-\cite{eom3}, scattering from a \emph{finite} number of apertures in a thick conducting screen is analyzed, but a numerical solution (matrix inversion) is required, with the matrix size increasing with the number of apertures. In this paper, we derive an analytical set of GSTCs for a metascreen of nonzero thickness, and demonstrate its validity when the thickness, aperture size and lattice constant are sufficiently small compared to a wavelength. A brief preliminary version of the present paper was presented in \cite{prelim}.

\section{Derivation of the GSTCs}

\begin{figure}[ht]
 \centering
 \scalebox{0.85}{\includegraphics*{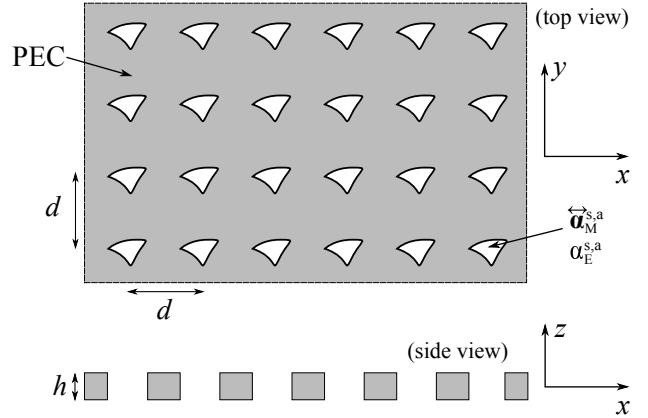}}
\caption{Top and side views of a metascreen consisting of a square array of identical apertures in a thick conducting screen.}
\label{f1}
\end{figure}

\begin{figure}[ht]
 \centering
 \scalebox{0.9}{\includegraphics*{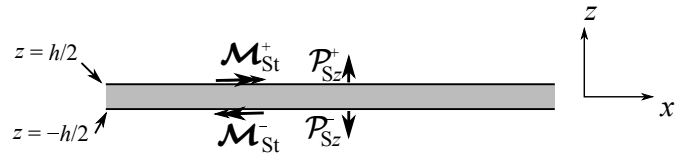}}
\caption{Side view of a thick metascreen showing equivalent surface polarization and magnetization densities.}
\label{f2}
\end{figure}

We will use the modified small-aperture coupling theory described in Appendix~\ref{apa} for screens of non-zero thickness to carry out a derivation analogous to that of \cite{ekel} for a screen of zero thickness. Our treatment will be limited to the case when the media on both sides of the screen are free space. Consider the metascreen shown in Fig.~\ref{f1}.
The apertures in the screen are arranged in a square array of lattice constant $d$. The screen is a perfect electric conductor (PEC) of thickness $h$, and each aperture of the array in isolation is described by its polarizabilities as described in Appendix~\ref{apa}. In the presence of a field, the effect of the apertures is to produce an additional field approximately equal to that produced by arrays of normal electric and tangential magnetic dipoles $\mathbf{p}_{\pm} = \mathbf{u}_z p_{z \pm}$ and $\mathbf{m}_{t \pm}$ located on the top and bottom faces $z = \pm h/2$ of a PEC screen with no holes (here $\mathbf{u}_d$ denotes a unit vector in the direction $d = x, y$ or $z$ in a cartesian coordinate system). These dipole arrays are in turn approximated by continuous distributions of surface polarization and magnetization densities:
\begin{equation}
 {\cal P}_{Sz}^{\pm} = N p_{z \pm} ; \qquad {\scb M}_{St}^{\pm} = N \mathbf{m}_{t \pm}
 \label{e1}
\end{equation}
where $N = 1/d^2$ is the density of apertures per unit area. The resulting situation is shown in Fig.~\ref{f2}.
From Appendix~A, the dipole moments can be expressed as
\begin{equation}
 \mathbf{p}_+ = - \epsilon_0 \alpha_E^s \left[ \mathbf{E}^{\rm sc} \right]_{z = -h/2}^{+h/2} - 2 \epsilon_0 \alpha_E^a \tilde{\mathbf{E}}^{\rm sc}_{\rm av}
 \label{pplus}
\end{equation}
\begin{equation}
 \mathbf{p}_- = + \epsilon_0 \alpha_E^s \left[ \mathbf{E}^{\rm sc} \right]_{z = -h/2}^{+h/2} - 2 \epsilon_0 \alpha_E^a \tilde{\mathbf{E}}^{\rm sc}_{\rm av}
 \label{pminus}
\end{equation}
and
\begin{equation}
 \mathbf{m}_+ = + \dyadic{\boldsymbol{\alpha}}_M^s \cdot \left[ \mathbf{H}^{\rm sc} \right]_{z = -h/2}^{+h/2} + 2 \dyadic{\boldsymbol{\alpha}}_M^a \cdot \tilde{\mathbf{H}}^{\rm sc}_{\rm av}
 \label{mplus}
\end{equation}
\begin{equation}
 \mathbf{m}_- = - \dyadic{\boldsymbol{\alpha}}_M^s \cdot \left[ \mathbf{H}^{\rm sc} \right]_{z = -h/2}^{+h/2} + 2 \dyadic{\boldsymbol{\alpha}}_M^a \cdot \tilde{\mathbf{H}}^{\rm sc}_{\rm av}
 \label{mminus}
\end{equation}
where the average of a field between the top and bottom sides of the screen is defined by
\begin{equation}
 \tilde{\mathbf{E}}_{\rm av} = \frac{1}{2} \left( \left. \mathbf{E} \right|_{z=h/2} + \left. \mathbf{E} \right|_{z=-h/2} \right)
\end{equation}
\begin{equation}
 \tilde{\mathbf{H}}_{\rm av} = \frac{1}{2} \left( \left. \mathbf{H} \right|_{z=h/2} + \left. \mathbf{H} \right|_{z=-h/2} \right)
\end{equation}
and $\alpha_E^{s,a}$ and $\dyadic{\boldsymbol{\alpha}}_M^{s,a}$ represent the symmetric and antisymmetric electric and magnetic aperture polarizabilities respectively, as defined in Appendix~\ref{apa}.

We now proceed using the same technique as in \cite{ekel}, referring the reader to that paper for the necessary details. We obtain expressions for the effective tangential electric field just above and just below the screen as:
\begin{eqnarray}
 \left. \mathbf{E} \right|_{z=h^+/2} \times \mathbf{u}_z & = & j \omega \mu_0 {\scb M}_{St}^+ - \nabla_t \left( \frac{{\cal P}_{Sz}^+}{\epsilon_0} \right) \times \mathbf{u}_z \label{etan} \\
 \left. \mathbf{E} \right|_{z=-h^+/2} \times \mathbf{u}_z & = & -j \omega \mu_0 {\scb M}_{St}^- + \nabla_t \left( \frac{{\cal P}_{Sz}^-}{\epsilon_0} \right) \times \mathbf{u}_z \nonumber
\end{eqnarray}
From (\ref{e1})-(\ref{mminus}) and the sum of the equations in (\ref{etan}), we get
\begin{eqnarray}
 \lefteqn{\tilde{\mathbf{E}}_{\rm av} \times \mathbf{u}_z = j \omega \mu_0 N \dyadic{\boldsymbol{\alpha}}_M^s \cdot \left[ \mathbf{H}^{\rm sc} \right]_{z = -h/2}^{+h/2} } \nonumber \\
 && \mbox{} + N \nabla_t \left( \alpha_E^s \left[ E^{\rm sc}_z \right]_{z = -h/2}^{+h/2} \right) \times \mathbf{u}_z \label{gstcp}
\end{eqnarray}
while from the difference of (\ref{etan}),
\begin{eqnarray}
 \lefteqn{\left[ \mathbf{E} \right]^{h/2}_{z=-h/2} \times \mathbf{u}_z = 4j \omega \mu_0 N \dyadic{\boldsymbol{\alpha}}_M^a \cdot \tilde{\mathbf{H}}^{\rm sc}_{\rm av} } \nonumber \\
 && \mbox{} + 4 N \nabla_t \left( \alpha_E^a \tilde{E}^{\rm sc}_{z, {\rm av}} \right) \times \mathbf{u}_z \label{gstcc}
\end{eqnarray}

The short-circuit fields $E^{\rm sc}_z$ and $\mathbf{H}^{\rm sc}$ are those that act on \emph{one} of the apertures when that aperture is filled with metal. As in \cite{ekel}, we have
\begin{eqnarray}
 \left[ \mathbf{H}^{\rm sc} \right]_{z = -h/2}^{+h/2} & = & \left[ \mathbf{H} \right]_{z = -h/2}^{+h/2} - 2 G(R) \left[ {\scb M}_{St}^+ - {\scb M}_{St}^- \right] \\
 & = & \left[ \mathbf{H} \right]_{z = -h/2}^{+h/2} - 4 G(R) N \dyadic{\boldsymbol{\alpha}}_M^s \cdot \left[ \mathbf{H}^{\rm sc} \right]_{z = -h/2}^{+h/2} \nonumber
\end{eqnarray}
or
\begin{eqnarray}
 \left[ H^{\rm sc}_x \right]_{z = -h/2}^{+h/2} & = & \frac{\left[ H_x \right]_{z = -h/2}^{+h/2}}{1 + 4 N \alpha_M^{s,xx} G(R)} \nonumber \\
 \left[ H^{\rm sc}_y \right]_{z = -h/2}^{+h/2} & = & \frac{\left[ H_y \right]_{z = -h/2}^{+h/2}}{1 + 4 N \alpha_M^{s,yy} G(R)}
\end{eqnarray}
where $\mathbf{H}$ is the effective field at the screen,
\begin{eqnarray}
 G(R) & = & - \frac{1}{4R} \left[ e^{-jk_0R} (1 - jk_0R) + 2jk_0R \right] \nonumber \\
 & = & - \frac{1}{4R} \left[ 1 + O(k_0^2 R^2) \right] \qquad (k_0R \ll 1),
\end{eqnarray}
$k_0 = \omega_0 \sqrt{\mu_0 \epsilon_0}$ is the wavenumber of free space, and
\begin{equation}
 R = \frac{2\pi d}{\sum'_{m,n} (m^2 + n^2)^{3/2}} \simeq 0.6956 d
 \label{esquare}
\end{equation}
is the exclusion radius for a square array of lattice constant $d$ (the prime on the double sum indicating that the term with $m=n=0$ is to be omitted). Likewise,
\begin{equation}
 \left[ E^{\rm sc}_z \right]_{z = -h/2}^{+h/2} = \frac{\left[ E_z \right]_{z = -h/2}^{+h/2}}{1 - 4 N \alpha_E^{s} F(R)}
\end{equation}
where
\begin{eqnarray}
 F(R) & = & \frac{1}{2R} e^{-jk_0R} (1 + jk_0R) \nonumber \\
 & = & \frac{1}{2R} \left[ 1 + O(k_0^2 R^2) \right] \qquad (k_0R \ll 1)
\end{eqnarray}
In a similar way, the average short-circuit fields are given by
\begin{eqnarray}
 \tilde{H}^{\rm sc}_{x,{\rm av}} & = & \frac{\tilde{H}_{x,{\rm av}}}{1 + 2 N \alpha_M^{a,xx} G(R)} \nonumber \\
 \tilde{H}^{\rm sc}_{y,{\rm av}} & = & \frac{\tilde{H}_{y,{\rm av}}}{1 + 2 N \alpha_M^{a,yy} G(R)} \label{tildeh}
\end{eqnarray}
and
\begin{equation}
 \tilde{E}^{\rm sc}_{z,{\rm av}} = \frac{\tilde{E}_{z,{\rm av}}}{1 - 2 N \alpha_E^{a} F(R)} \label{tildee}
\end{equation}

Substituting these results into (\ref{gstcp}) and approximating
\begin{equation}
 F(R) \simeq \frac{1}{2R} ; \qquad G(R) \simeq - \frac{1}{4R}
\end{equation}
which is valid so long as $k_0 R \ll 1$, we get a first boundary condition for the metascreen:
\begin{eqnarray}
 \tilde{\mathbf{E}}_{\rm av} \times \mathbf{u}_z & = & j \omega \mu_{0} \tilde{\dyadictall{\boldsymbol{\pi}}}_{MS}^t \cdot \left[ \mathbf{H}_t \right]_{z=-h/2}^{h/2} \nonumber \\
 && - \nabla_t \left\{ \tilde{\pi}_{ES}^{zz} \left[ E_z \right]_{z=-h/2}^{h/2} \right\} \times \mathbf{u}_z \label{egstc}
\end{eqnarray}
where
\begin{eqnarray}
 \tilde{\pi}_{ES}^{zz} & = & - \frac{N \alpha_E^s}{1 - \frac{2N}{R} \alpha_E^s} \label{tilpi} \\
 \tilde{\dyadictall{\boldsymbol{\pi}}}_{MS}^t & = & \mathbf{u}_x \mathbf{u}_x \frac{N \alpha_M^{s,xx}}{1 - \frac{N}{R} \alpha_M^{s,xx}} + \mathbf{u}_y \mathbf{u}_y \frac{N \alpha_M^{s,yy}}{1 - \frac{N}{R} \alpha_M^{s,yy}} \nonumber
\end{eqnarray}
are electric and magnetic surface porosities of the metascreen (relative to the planes $z = \pm h/2$), respectively. The choice of sign for $\tilde{\pi}_{ES}^{zz}$ is the same as that used in \cite{ekel} for a metascreen of vanishing thickness. A second boundary condition is obtained in a similar way from (\ref{gstcc}):
\begin{equation}
 \left[ \mathbf{E} \right]_{z = -h/2}^{h/2} \times \mathbf{u}_z = j \omega \mu_{0} \tilde{\dyadictall{\boldsymbol{\chi}}}_{MS}^t \cdot \tilde{\mathbf{H}}_{\rm av} + \mathbf{u}_z \times \nabla_t \left( \tilde{\chi}_{ES}^{zz} \tilde{E}_{z,{\rm av}} \right)
\label{sgstc}
\end{equation}
where
\begin{eqnarray}
 \tilde{\chi}_{ES}^{zz} & = & -\frac{4N \alpha_E^a}{1 - \frac{N}{R} \alpha_E^a} \label{tilchi} \\
 \tilde{\dyadictall{\boldsymbol{\chi}}}_{MS}^t & = & \mathbf{u}_x \mathbf{u}_x \frac{4N \alpha_M^{a,xx}}{1 - \frac{N}{2R} \alpha_M^{a,xx}} + \mathbf{u}_y \mathbf{u}_y \frac{4N \alpha_M^{a,yy}}{1 - \frac{N}{2R} \alpha_M^{a,yy}} \nonumber
\end{eqnarray}
are electric and magnetic surface susceptibilities of the metascreen, again relative to the top and bottom surfaces of the metal screen. This condition is of the same form as equation (81) of \cite{kmh}, which applies to a metafilm.

The reason we have used tildes in the notation for the fields (\ref{tildeh})-(\ref{tildee}) and the surface porosities and susceptibilities in (\ref{tilpi}) and (\ref{tilchi}) is that here, these quantities describe relationships between the fields at $z = \pm h/2$. However, according to the general definition of sheet transition conditions given by Senior and Volakis \cite{senior}, they should apply to an equivalent sheet of zero thickness, like those in \cite{ekel} and \cite{kmh}, where the effective fields in the GSTCs are evaluated at $z = 0^{\pm}$ (extrapolated to these positions if necessary). If $h$ is small compared to a wavelength, let us seek a set of GSTCs that conform to this requirement. To do this, we use Taylor series expansions:
\begin{equation}
 \left. \mathbf{E}_t \right|_{z = \pm h/2} \simeq \left. \mathbf{E}_t \right|_{z = 0^{\pm}} \pm \frac{h}{2} \left. \frac{\partial \mathbf{E}_t}{\partial z} \right|_{z = 0^{\pm}}
 \label{taylor}
\end{equation}
From the transverse components of Faraday's law, we have
\begin{equation}
 - \mathbf{u}_z \times \nabla_t E_z + \mathbf{u}_z \times \frac{\partial \mathbf{E}_t}{\partial z} = - j \omega \mu_0 \mathbf{H}_t
 \label{faradayt}
\end{equation}
Using (\ref{taylor}), and (\ref{faradayt}) at $z=0^{\pm}$, in (\ref{egstc}) and (\ref{sgstc}), and neglecting second order terms proportional to $h \pi_{(e,m)s}$ or $h \chi_{(e,m)s}$ that appear on the right sides, we obtain true GSTCs that apply at an equivalent surface of zero thickness:
\begin{equation}
 \mathbf{E}_{\rm av} \times \mathbf{u}_z = j \omega \mu_{0} \dyadictall{\boldsymbol{\pi}}_{MS}^t \cdot \left[ \mathbf{H}_t \right]_{z=0^-}^{0^+} - \nabla_t \left\{ \pi_{ES}^{zz} \left[ E_z \right]_{z=0^-}^{0^+} \right\} \times \mathbf{u}_z \label{egstctrue}
\end{equation}
and
\begin{equation}
 \left[ \mathbf{E} \right]_{z = 0^-}^{0^+} \times \mathbf{u}_z = j \omega \mu_{0} \dyadictall{\boldsymbol{\chi}}_{MS}^t \cdot \mathbf{H}_{\rm av} - \nabla_t \left( \chi_{ES}^{zz} E_{z,{\rm av}} \right) \times \mathbf{u}_z
\label{sgstctrue}
\end{equation}
where
\begin{equation}
 \mathbf{E}_{\rm av} = \frac{1}{2} \left( \left. \mathbf{E} \right|_{z=0^+} + \left. \mathbf{E} \right|_{z=0^-} \right)
\end{equation}
\begin{equation}
 \tilde{\mathbf{H}}_{\rm av} = \frac{1}{2} \left( \left. \mathbf{H} \right|_{z=0^+} + \left. \mathbf{H} \right|_{z=0^-} \right)
\end{equation}
are the average fields across the zero-thickness equivalent surface at $z=0$, and
\begin{eqnarray}
 \pi_{ES}^{zz} & = & \tilde{\pi}_{ES}^{zz} + \frac{h}{4} \nonumber \\
 \dyadictall{\boldsymbol{\pi}}_{MS}^t & = & \tilde{\dyadictall{\boldsymbol{\pi}}}_{MS}^t - \frac{h}{4} \left( \mathbf{u}_x \mathbf{u}_x + \mathbf{u}_y \mathbf{u}_y \right) \nonumber \\
 \chi_{ES}^{zz} & = & \tilde{\chi}_{ES}^{zz} + h \nonumber \\
 \dyadictall{\boldsymbol{\chi}}_{MS}^t & = & \tilde{\dyadictall{\boldsymbol{\chi}}}_{MS}^t - h \left( \mathbf{u}_x \mathbf{u}_x + \mathbf{u}_y \mathbf{u}_y \right) \label{pichi}
\end{eqnarray}

Equation (\ref{sgstctrue}) replaces the condition that tangential $\mathbf{E}$ be continuous at $z=0$, which holds for a metascreen of zero thickness \cite{ekel}. Equation (\ref{egstctrue}) has the same form as the GSTC obtained in \cite{ekel} in that case, but with different values of the surface porosities. These forms have also been obtained using the method of multiple-scale homogenization \cite{hkscreen}, wherein the surface porosities and susceptibilities are found from solutions of certain electrostatic and magnetostatic field problems. This technique gives results not limited by the assumption of a dipole-interaction model, but in general requires numerical solutions of the relevant static field problems, whereas the method of the present paper gives closed-form analytical expressions for the surface parameters. The GSTCs derived here also resemble equations (147) and (149) of \cite{hkdrs}, which were obtained for a wire grating.

We may convert one of our GSTCs into a somewhat different form by expressing the surface current density as
\begin{equation}
 \mathbf{J}_S = \mathbf{u}_z \times \left[ \mathbf{H}_t \right]_{z=0^-}^{0^+}
\end{equation}
and using the result
\begin{equation}
 E_z = - \frac{1}{j\omega \epsilon_0} \nabla_t \cdot \left( \mathbf{u}_z \times \mathbf{H}_t \right)
\end{equation}
that follows from Amp\`{e}re's law. Then (\ref{egstctrue}) can be expressed as
\begin{equation}
 \mathbf{E}_{t,{\rm av}} = j \dyadictall{\mathbf{X}}_{ms} \cdot \mathbf{J}_S + \frac{1}{j\omega \epsilon_0} \nabla_t \left( \pi_{ES}^{zz} \nabla_t \cdot \mathbf{J}_S \right) \label{egstckontform}
\end{equation}
where
\begin{eqnarray}
 \lefteqn{\dyadictall{\mathbf{X}}_{ms} = \omega \mu_0 \left( \mathbf{u}_x \mathbf{u}_x \pi_{MS}^{yy} \right.} \\
 && \left. \mbox{} - \mathbf{u}_x \mathbf{u}_y \pi_{MS}^{yx} - \mathbf{u}_y \mathbf{u}_x \pi_{MS}^{xy} + \mathbf{u}_y \mathbf{u}_y \pi_{MS}^{xx} \right) \nonumber
\end{eqnarray}
is the dyadic surface reactance of the metascreen. Equation (\ref{egstckontform}) has the form of the boundary condition obtained by Kontorovich and his colleagues \cite{kont2}-\cite{kont4} for a thin-wire mesh. There is no analog of (\ref{sgstctrue}) in the Kontorovich model; it assumes that tangential $\mathbf{E}$ is continuous, as is the case for a metascreen of zero thickness \cite{ekel}.

\section{Equivalent Circuit}

An equivalent circuit for a thick metascreen can be obtained under certain conditions. Suppose that the field has no variation in the $y$-direction ($\partial/\partial y \equiv 0$) and that all fields vary with $x$ as $e^{-j k_x x}$. Suppose moreover that the magnetic porosity and susceptibiity dyadics are diagonal: $\dyadic{\boldsymbol{\pi}}_{MS}^t = \mathbf{u}_x \mathbf{u}_x \pi_{MS}^{xx} + \mathbf{u}_y \mathbf{u}_y \pi_{MS}^{yy}$ and $\dyadic{\boldsymbol{\chi}}_{MS}^t = \mathbf{u}_x \mathbf{u}_x \chi_{MS}^{xx} + \mathbf{u}_y \mathbf{u}_y \chi_{MS}^{yy}$. Then it is readily shown that the field can be written as the superposition of a TE part (consisting of the field components $E_y$, $H_x$ and $H_z$ only) and a TM part (consisting of the field components $H_y$, $E_x$ and $E_z$ only), no conversion occurring between these two polarizations.

\begin{figure}[ht]
 \centering
 \scalebox{1.2}{\includegraphics*{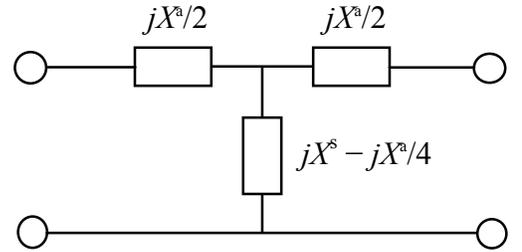}}
\caption{Equivalent circuit of a thick metascreen located at $z=0$.}
\label{feqckt}
\end{figure}

For the TE field, let $E_y \rightarrow V$ and $H_x \rightarrow -I$. Inserting these into (\ref{egstctrue}) and (\ref{sgstctrue}), we find that the metascreen can be represented by the equivalent circuit of Figure~\ref{feqckt} placed at $z=0$, wherein
\begin{equation}
 X^s_{\rm TE} = \omega \mu_{0} \pi_{MS}^{xx} ; \qquad X^a_{\rm TE} = \omega \mu_{0} \chi_{MS}^{xx}
 \label{eqTE}
\end{equation}
are respectively the symmetric and antisymmetric TE reactances of the metascreen.
Likewise, for the TM field let $E_x \rightarrow V$ and $H_y \rightarrow I$. By Amp\`{e}re's law and the assumptions above about the $x$- and $y$-dependences of the field, we have $E_z = -(k_x/\omega \epsilon_0) H_y$. Once more the GSTCs (\ref{egstctrue}) and (\ref{sgstctrue}) are represented by the network of Figure~\ref{feqckt}, with now
\begin{eqnarray}
 X^s_{\rm TM} & = & \omega \mu_{0} \left( \pi_{MS}^{yy} + \frac{k_x^2}{k_0^2} \pi_{ES}^{zz} \right) \nonumber \\
 X^a_{\rm TM} & = & \omega \mu_{0} \left( \chi_{MS}^{yy} - \frac{k_x^2}{k_0^2} \chi_{ES}^{zz} \right) \label{eqTM}
\end{eqnarray}

\section{Plane Wave Reflection and Transmission}

In this section, we will apply the GSTCs obtained above to the determination of the reflection and transmission coefficients of a plane wave incident on a thick metascreen. The procedure is very similar to that used in \cite{ekel}, and we will omit much of the detail.
\begin{figure}[ht]
 \centering
 \scalebox{0.5}{\includegraphics*{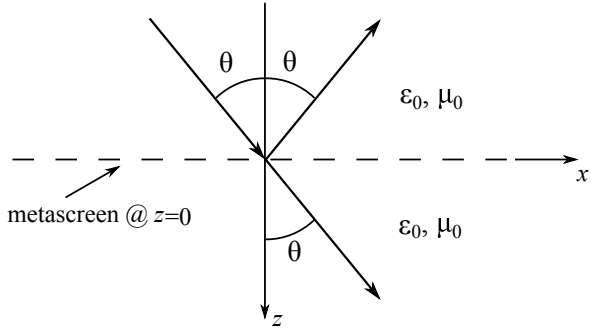}}
\caption{Plane wave incident at a metascreen.}
\label{f4}
\end{figure}

If a TE (perpendicular) polarized plane wave is incident at an angle $\theta$ to the $z$-axis as shown in Figure~\ref{f4}, the electric field $\mathbf{E} = \mathbf{u}_y E_y$ is given by
\begin{equation}
\begin{array}{l}
 E_y = e^{-jk_0 x \sin \theta} \left[ e^{-jk_0 z \cos \theta} + \Gamma_{\rm TE} e^{jk_0 z \cos \theta} \right] \quad \mbox{\rm ($z<0$)} \\
 \quad \ \, = e^{-jk_0 x \sin \theta} T_{\rm TE} e^{-jk_0 z \cos \theta} \quad \mbox{\rm ($z>0$)}
 \end{array}
\end{equation}
where $\Gamma_{\rm TE}$ is the reflection coefficient and $T_{\rm TE}$ is the transmission coefficient. The magnetic field is obtained from Faraday's law, so enforcing the GSTCs (\ref{egstctrue}) and (\ref{sgstctrue}) at $z=0$ in the usual way leads to:
\begin{equation}
 \Gamma_{\rm TE} = - 1 + j \frac{2 X^s_{\rm TE} \frac{\cos \theta}{\zeta_0}}{1 +2j X^s_{\rm TE} \frac{\cos \theta}{\zeta_0}} + j \frac{X^a_{\rm TE} \frac{\cos \theta}{2 \zeta_0}}{1 + j X^a_{\rm TE} \frac{\cos \theta}{2 \zeta_0}}
 \label{gammaTE}
\end{equation}
and
\begin{equation}
 T_{\rm TE} = j \frac{2 X^s_{\rm TE} \frac{\cos \theta}{\zeta_0}}{1 +2j X^s_{\rm TE} \frac{\cos \theta}{\zeta_0}} - j \frac{X^a_{\rm TE} \frac{\cos \theta}{2 \zeta_0}}{1 + j X^a_{\rm TE} \frac{\cos \theta}{2 \zeta_0}}
 \label{TTE}
\end{equation}
where $\zeta_0 = \sqrt{\mu_0/\epsilon_0}$ is the wave impedance of free space. These formulas could also have been obtained by using the equivalent circuit (\ref{eqTE}) placed at $z=0$ between two sections of transmission line with characteristic impedance $\zeta_0/\cos \theta$, using $k_x = k_0 \sin \theta$. We observe that for an unperforated PEC screen located at $z = - h/2$, we have $\tilde{\pi}_{MS}^{xx}$ and $\tilde{\chi}_{MS}^{xx} \rightarrow 0$, so we obtain $T_{\rm TE} = 0$ and
\begin{equation}
 \Gamma_{\rm TE} = - e^{2j \tan^{-1} \left( k_0 h \cos \theta /2 \right)} \simeq - e^{j k_0 h \cos \theta } ,
\end{equation}
as expected on physical grounds.

Reflection and transmission coefficients for the TM polarization are obtained in a similar way:
\begin{equation}
 \Gamma_{\rm TM} = - 1 + j \frac{ \frac{2 X^s_{\rm TM}}{\zeta_0 \cos \theta }}{1 +2j \frac{X^s_{\rm TM}}{\zeta_0 \cos \theta }} + j \frac{\frac{X^a_{\rm TM}}{2\zeta_0 \cos \theta}}{1 + j \frac{X^a_{\rm TM}}{2\zeta_0 \cos \theta }}
 \label{gammaTM}
\end{equation}
and
\begin{equation}
 T_{\rm TM} = j \frac{\frac{2 X^s_{\rm TM}}{\zeta_0 \cos \theta }}{1 + 2j \frac{X^s_{\rm TM}}{\zeta_0 \cos \theta }} - j \frac{\frac{X^a_{\rm TM}}{2 \zeta_0 \cos \theta }}{1 + j \frac{X^a_{\rm TM}}{2 \zeta_0 \cos \theta }}
 \label{TTM}
\end{equation}
which are also obtainable by using the equivalent circuit (\ref{eqTM}) placed at $z=0$ between two sections of transmission line with characteristic impedance $\zeta_0 \cos \theta$.

An interesting special case is obtained for TE polarization if $X^s_{\rm TE} = - X^a_{\rm TE}/4$, that is, $\pi_{MS}^{xx} = -\chi_{MS}^{xx}/4$. Under this condition, the phases of $\Gamma_{\rm TE}$ and $T_{\rm TE}$ are independent of frequency (at least under the low-frequency approximation for which our GSTCs are valid)---in fact,
\begin{eqnarray}
 \Gamma_{\rm TE} & = & - \frac{1 - \left( \frac{X^a_{\rm TE} \cos \theta}{2\zeta_0} \right)^2}{1 + \left( \frac{X^a_{\rm TE} \cos \theta}{2\zeta_0} \right)^2} \nonumber \\
  T_{\rm TE} & = & - 2j \frac{\frac{X^a_{\rm TE} \cos \theta}{2\zeta_0} }{1 + \left( \frac{X^a_{\rm TE} \cos \theta}{2\zeta_0} \right)^2}
\end{eqnarray}
This frequency-independent behavior of the phases occurs no matter the angle of incidence. The same behavior will occur in the TM polarization if $X^s_{\rm TM} = - X^a_{\rm TM}/4$, but the presence of $k_x = k_0 \sin \theta$ in (\ref{eqTM}) means that it can happen only for a single angle of incidence given fixed aperture size and screen thickness. A similar phenomenon has been previously observed for a grating of parallel wires \cite{sivov}, \cite{chuprin}, and is explained by the compensation of the phase shift due to the thickness of the metascreen with that of the distortion of the local fields in the neighborhood of the apertures.

\begin{figure}[ht!]
 \centering
 \scalebox{0.91}{\includegraphics*{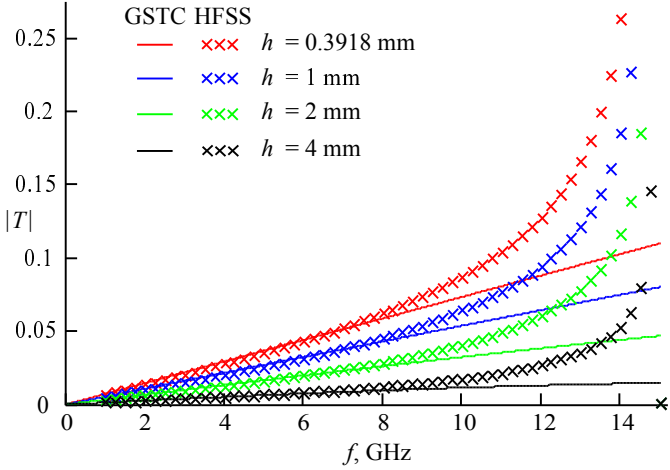}}
\caption{Magnitude of the transmission coefficient of a normally incident plane wave at a metascreen with a square array of circular holes of radius $r_0 = 4$ mm with lattice constant $d = 20$ mm.}
\label{f5}
\end{figure}

\begin{figure}[ht!]
 \centering
 \scalebox{0.91}{\includegraphics*{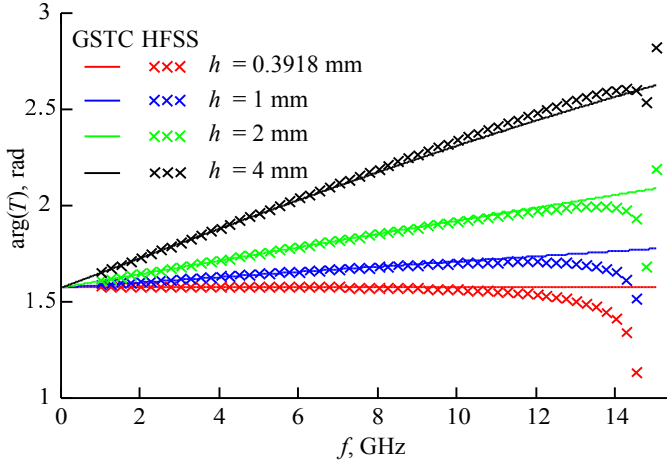}}
\caption{Phase of the transmission coefficient of a normally incident plane wave at a metascreen with a square array of circular holes of radius $r_0 = 4$ mm with lattice constant $d = 20$ mm.}
\label{f6}
\end{figure}

\begin{figure}[ht!]
 \centering
 \scalebox{0.9}{\includegraphics*{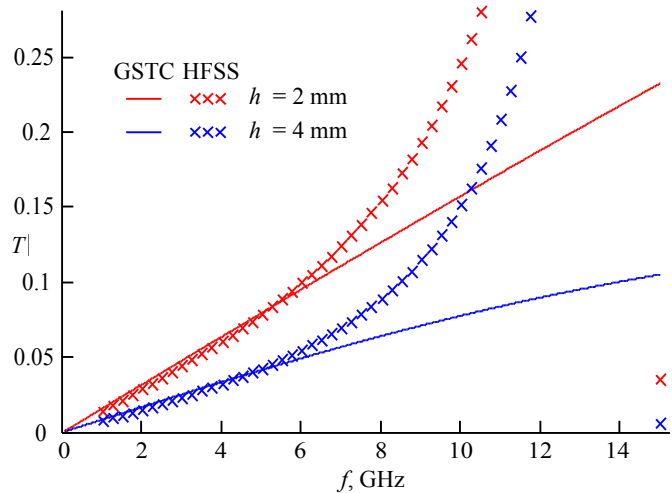}}
\caption{Magnitude of the transmission coefficient of a normally incident plane wave at a metascreen with a square array of circular holes of radius $r_0 = 6$ mm with lattice constant $d = 20$ mm.}
\label{f7}
\end{figure}

\begin{figure}[ht!]
 \centering
 \scalebox{0.9}{\includegraphics*{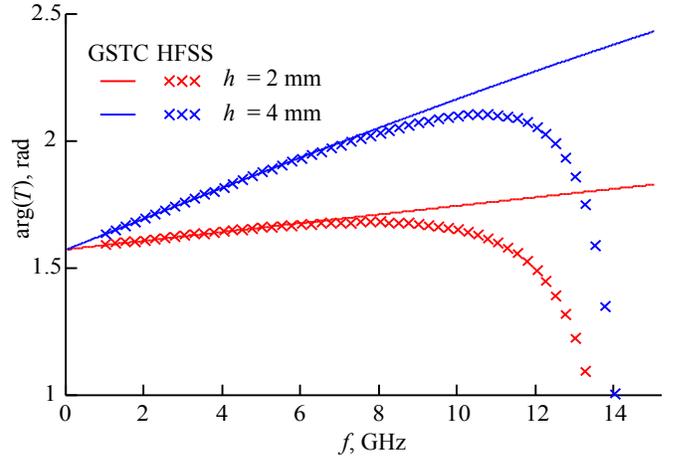}}
\caption{Phase of the transmission coefficient of a normally incident plane wave at a metascreen with a square array of circular holes of radius $r_0 = 6$ mm with lattice constant $d = 20$ mm.}
\label{f8}
\end{figure}

To illustrate the accuracy of the GSTC representation, we compare the transmission coefficient of a normally incident plane wave from a metascreen as computed from (\ref{TTE}) or (\ref{TTM}) with results of a full-wave finite-element simulation (HFSS from ANSYS). We have chosen the apertures to be circles of radius $r_0$ and a lattice constant $d = 20$ mm, which is equal to a half wavelength at $f_{\lambda/2} = 7.5$ GHz. In Figures~\ref{f5} and \ref{f6}, we see that the magnitude and phase of the transmission coefficient show good agreement between the GSTC prediction and numerical results for a hole radius $r_0 = 4$ mm and a variety of screen thicknesses, up to well above $f_{\lambda/2}$. The thickness $h = 0.3918$ mm was chosen to give frequency-independent phase, and we see that this is indeed well realized by the full-wave simulation up to nearly 10 GHz.
For holes of larger radius $r_0 = 6$ mm, we can see from Figures~\ref{f7} and \ref{f8} that good agreement between GSTC and full-wave results does not extend to such high frequencies, but is still quite good up to $f_{\lambda/2}$, especially for the phase.
The poorer agreement is not too surprising, given that the derivation of our expressions for surface porosities and susceptibilities is based on the dipole interaction approximation, which in principle requires that $2r_0$ be small compared to $d$. In such cases, the method of \cite{hkscreen} should be used. In all the situations we have examined, good agreement for the phase extends to significantly higher frequencies than for the magnitude.

\section{Conclusion}

In this paper, we have obtained generalized sheet transition conditions for describing the interaction of electromagnetic waves with periodically perforated conducting screens. In most other related work (e.~g., \cite{kaplun}-\cite{garcia}), full-wave formulations have been used to obtain only numerical results for plane-wave reflection and transmission. Our results have been derived using the dipole-interaction approximation (analogous to what is used to obtain the Clausius-Mossotti-Lorentz-Lorenz approximation for effective dielectric constant). Because of that, our formulas are only certain to be valid if the dimensions of the aperture are small compared to the lattice constant $d$. Nevertheless, the comparisons with full-wave results indicate that, at least for circular apertures, good accuracy can be obtained even when the aperture size is greater than half the lattice constant, and up to at least a frequency where that lattice constant is 40\% of a wavelength.

We should note that, for the case of normal incidence at a square array of square apertures with side $a$, Chen \cite{chen} has obtained results that have the same form as (\ref{gammaTE})-(\ref{TTE}) or (\ref{gammaTM})-(\ref{TTM}) [his equations (6)-(7)]. Lee \emph{et al.} \cite{lee} have given a corrected version of Chen's result for the transmission coefficient [equation (14a)] that is identical in form to ours.\footnote{
There is an error in the first line of Lee's equation (14a): in the denominator, the term $-Z \tan (\Gamma \tau/2)$ should be $+Z \tan (\Gamma \tau/2)$.}
Equating \cite[eqn. (14a)]{lee} to (\ref{TTE}) or (\ref{TTM}) for $\theta = 0$ allows expressions for $\pi_{MS}^{xx} = \pi_{MS}^{yy} \equiv \pi_{MS}^t$ and $\chi_{MS}^{xx} = \chi_{MS}^{yy} \equiv \chi_{MS}^t$ to be inferred. These are quite cumbersome in their general forms, but if a small argument expansion is carried out for $k_0 d \ll 1$ and only the terms of order $k_0d$ and $(k_0d)^2$ are retained, the surface parameters become in our notation
\begin{eqnarray}
 \pi_{MS}^t & = & \frac{\pi_{MS}^{0t}}{\displaystyle 1 + 2 \left( \frac{\pi d}{2a} \right)^3 \frac{\pi_{MS}^{0t}}{d} \tanh \frac{\pi h}{2a}} - \frac{h}{4} \nonumber \\
 \chi_{MS}^t & = & \frac{4\pi_{MS}^{0t}}{\displaystyle 1 + 2 \left( \frac{\pi d}{2a} \right)^3 \frac{\pi_{MS}^{0t}}{d} \coth \frac{\pi h}{2a}} - h \label{square}
\end{eqnarray}
where $\pi_{MS}^{0t}$ is the magnetic surface porosity for the screen of zero thickness.\footnote{
A formula similar to (\ref{square}) can be inferred from a result in \cite{matsu}, but its parameter dependence is different, and seems to predict an increase in $\pi_{MS}^t$ with screen thickness, rather than a decrease (which is what would be expected on physical grounds).}
Rather than use the formula for $\pi_{MS}^{0t}$ that can be inferred from \cite{chen} and \cite{lee}, we choose instead to use that of \cite{ekel}, which was shown to be more accurate over the entire range of $a/d$:
\begin{equation}
 \frac{\pi_{MS}^{0t}}{d} = \frac{\ln \sec \frac{\pi a}{2d}}{2\pi} \left[ C_2 \frac{a}{d} + \left( 1 - C_2 \right) \frac{a^2}{d^2} + \frac{\sin \left( \pi \frac{a^2}{d^2} \right)}{25} \right]
 \label{apc1}
\end{equation}
where
\begin{equation}
 C_2 = \frac{32}{9\pi \ln (1 + \sqrt{2})} = 1.2841 \ldots
\end{equation}
The formula for $T$ from \cite{chen} and \cite{lee} is only claimed to be valid if $a \geq 0.7 d$; clearly (\ref{square}) cannot be expected to reduce to our results, which have been derived on the assumption that $a$ is small compared to $d$. A similar procedure starting from \cite[eqns. (11)-(12)]{chen} for a square array of circular holes of radius $r_0$ leads to
\begin{eqnarray}
 \pi_{MS}^t & = & \frac{\pi_{MS}^{0t}}{\displaystyle 1 + \frac{2.4\,j'_{11}}{\pi} \left( \frac{d}{r_0} \right)^3 \frac{\pi_{MS}^{0t}}{d} \tanh \frac{j'_{11} h}{2r_0}} - \frac{h}{4} \nonumber \\
 \chi_{MS}^t & = & \frac{4\pi_{MS}^{0t}}{\displaystyle 1 + \frac{2.4\,j'_{11}}{\pi} \left( \frac{d}{r_0} \right)^3 \frac{\pi_{MS}^{0t}}{d} \coth \frac{j'_{11} h}{2r_0}} - h \label{circle}
\end{eqnarray}
where $j'_{11} = 1.841 \ldots$ is the first root of the Bessel function derivative $J'_1(x)$ and now the best known approximation for the zero-thickness porosity is \cite{ekel}:
\begin{equation}
 \pi_{MS}^{0t} = \frac{4 r_0^3}{3d^2} \frac{1}{1 - \frac{4 r_0^3}{3Rd^2}} ; \qquad  R \simeq 0.6956 d
 \label{circsmall}
\end{equation}

All of the foregoing formulas were derived without the restriction that the aperture size be small compared to the lattice constant. This suggests that it should be possible to use the GSTC model for a metascreen outside of the constraint that the holes be small compared to the lattice constant, provided that suitable formulas for the surface porosities and surface susceptibilities can be found. Indeed, the multiple-scales homogenization method used in \cite{hkscreen} has shown that GSTCs of the same form as those derived in the present paper are applicable to metascreens of very general shape, provided only that the lattice constant of the metascreen is small enough compared to a wavelength. Even that restriction can probably be relaxed, as shown in \cite{medina} where frequency-dependent equivalent circuit parameters were obtained by fitting to simulated scattering data and used to predict the metascreen behavior well above the first resonant frequency.

\appendices

\section{Aperture Polarizabilities of a Thick PEC Screen}
\label{apa}

Many authors have proposed modifications of the Bethe small-aperture theory (which describes electromagnetic coupling between two regions of space separated by a perfectly conducting screen of zero thickness containing an electrically small aperture) so as to be applicable to apertures in a screen of nonzero thickness \cite{marc}-\cite{nik}, \cite{eom3}. In this appendix, we will summarize this work, and present a form more adapted to the needs of the present paper.

\subsection*{\ref{apa}.1 Circular aperture}

\begin{figure}[ht]
 \centering
 \scalebox{0.85}{\includegraphics*{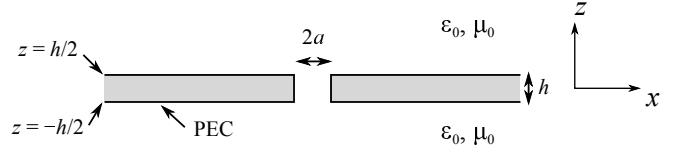}}
\caption{Aperture in a thick PEC screen.}
\label{fa1}
\end{figure}

\begin{figure}[ht]
 \centering
 \scalebox{0.85}{\includegraphics*{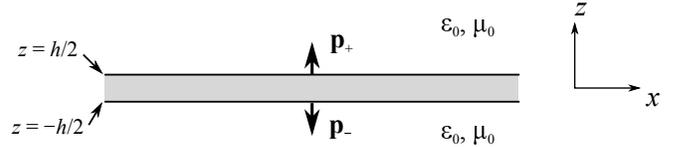}}
\caption{Equivalent electric dipoles for an aperture in a thick PEC screen.}
\label{fa2}
\end{figure}

To start with, we consider a circular aperture of radius $r_0$ in a perfectly conducting sheet that lies between $z=-h/2$ and $+h/2$ as shown in Figure~\ref{fa1}.
For simplicity we assume that the remainder of space is vacuum. Some given sources located outside the aperture produce a field $\mathbf{E}$, $\mathbf{H}$ in the presence of this punctured screen. According to \cite{gluck}, the effect of the aperture can be replaced by that of dipoles placed at $z = \pm h/2$, with the aperture filled by more PEC as shown in Figure~\ref{fa2} for the electric dipoles.
We can express the electric dipole moments as
\begin{equation}
 \mathbf{p}_+ = - \epsilon_0 \left( \alpha_E^s + \alpha_E^a \right) \left. \mathbf{E}^{\rm sc} \right|_{z = h/2} + \epsilon_0 \left( \alpha_E^s - \alpha_E^a \right) \left. \mathbf{E}^{\rm sc} \right|_{z = - h/2} 
\end{equation}
\begin{equation}
 \mathbf{p}_- = - \epsilon_0 \left( \alpha_E^s + \alpha_E^a \right) \left. \mathbf{E}^{\rm sc} \right|_{z = - h/2} + \epsilon_0 \left( \alpha_E^s - \alpha_E^a \right) \left. \mathbf{E}^{\rm sc} \right|_{z = h/2}
\end{equation}
where $\mathbf{E}^{\rm sc}$ is the electric field that would be produced by the given sources when the aperture filled with a PEC, $\alpha_E^s$ is what we will call the symmetric electric polarizability of the aperture, and $\alpha_E^a$ is its antisymmetric electric polarizability (we use different notations from those in \cite{gluck}). If the thickness $h \rightarrow 0$, then $\alpha_E^a \rightarrow 0$ and $\alpha_E^a \rightarrow \alpha_{E}^0 = 2r_0^3/3$ (the value of the electric polarizability when the screen thickness is zero), and we recover the result of ordinary small aperture theory.

More precisely, we can obtain from \cite{gluck} expressions for the limiting values of the polarizabilities:
\begin{equation}
\left.
 \begin{array}{rcl}
  \alpha_E^s & \simeq & \alpha_{E}^0 - \frac{hr_0^2}{\pi} \left( \ln \frac{r_0}{h} + 1.88 \right) \\
  \alpha_E^a & \simeq & \frac{\pi h r_0^2}{4}
 \end{array}
\right\} 
 \quad \mbox{\rm ($h/r_0 \ll 1$)}
 \label{aE}
\end{equation}
and
\begin{equation}
\left.
 \begin{array}{rcl}
  \alpha_E^s & \simeq & \alpha_{E}^{\infty} + 0.412 \alpha_{E}^0 e^{-2.405 \frac{h}{r_0}} \\
  \alpha_E^a & \simeq & \alpha_{E}^{\infty} - 0.412 \alpha_{E}^0 e^{-2.405 \frac{h}{r_0}}
 \end{array}
\right\} 
 \quad \mbox{\rm ($h/r_0 \gg 1$)}
 \label{aEl}
\end{equation}
where
\begin{equation}
 \alpha_{E}^{\infty} \simeq 0.42923 \alpha_E^0 = 0.28615 r_0^3
\end{equation}
is the limiting value of $\alpha_E^{s,a}$ as $h/a \rightarrow \infty$ (see also \cite{shap}\footnote{The values given therein have been converted to agree with our definitions.}).

In a similar way, there are equivalent magnetic dipoles at $z = \pm h/2$, which are given by
\begin{equation}
 \mathbf{m}_+ = \left( \dyadic{\boldsymbol{\alpha}}_M^s + \dyadic{\boldsymbol{\alpha}}_M^a \right) \cdot \left. \mathbf{H}^{\rm sc} \right|_{z = h/2} - \left( \dyadic{\boldsymbol{\alpha}}_M^s - \dyadic{\boldsymbol{\alpha}}_M^a \right) \cdot \left. \mathbf{H}^{\rm sc} \right|_{z = -h/2}
\end{equation}
\begin{equation}
 \mathbf{m}_- = \left( \dyadic{\boldsymbol{\alpha}}_M^s + \dyadic{\boldsymbol{\alpha}}_M^a \right) \cdot \left. \mathbf{H}^{\rm sc} \right|_{z = -h/2} - \left( \dyadic{\boldsymbol{\alpha}}_M^s - \dyadic{\boldsymbol{\alpha}}_M^a \right) \cdot \left. \mathbf{H}^{\rm sc} \right|_{z = h/2}
\end{equation}
where the dyadic magnetic polarizabilities are diagonal: $\dyadic{\boldsymbol{\alpha}}_M^{(s,a)} = \alpha_M^{(s,a),tt} (\mathbf{u}_x \mathbf{u}_x + \mathbf{u}_y \mathbf{u}_y)$, and from \cite{gluck} we have
\begin{equation}
\left.
 \begin{array}{rcl}
  \alpha_M^{s,tt} & \simeq & \alpha_{M}^0 - \frac{2hr_0^2}{\pi} \left( \ln \frac{r_0}{h} + 1.88 \right) \\
  \alpha_M^{a,tt} & \simeq & \frac{\pi h r_0^2}{4}
 \end{array}
\right\} 
 \quad \mbox{\rm ($h/r_0 \ll 1$)}
 \label{aM}
\end{equation}
and\footnote{There appears to be a misprint in equation (59) of \cite{gluck}: the constant 0.716 should probably be 0.167, which gives much better agreement with the results in Table I of that paper when $h/r_0 \geq 0.3$.}
\begin{equation}
\left.
 \begin{array}{rcl}
  \alpha_M^{s,tt} & \simeq & \alpha_{M}^{\infty} + 0.423 \alpha_{M}^0 e^{-1.841 \frac{h}{r_0}} \\
  \alpha_M^{a,tt} & \simeq & \alpha_{M}^{\infty} - 0.423 \alpha_{M}^0 e^{-1.841 \frac{h}{r_0}}
 \end{array}
\right\} 
 \quad \mbox{\rm ($h/r_0 \gg 1$)}
 \label{aMl}
\end{equation}
where $\alpha_{M}^0 = 4r_0^3/3$ is the magnetic polarizability when the screen thickness is zero, and
\begin{equation}
 \alpha_{M}^{\infty} \simeq 0.35494 \alpha_{M}^0 = 0.47325 r_0^3
 \label{aMinf0}
\end{equation}
is the limiting value of $\alpha_M^{s,a}$ as $h/r_0 \rightarrow \infty$ (see also \cite{smythe} and \cite{shap}).

There is a physical significance to the exponentially small terms in these expressions---they represent the effect of a cutoff mode in a section of circular waveguide extending from one face of the aperture to the other. For the electric polarizabilities, this is the TM$_{01}$ mode, while for the magnetic polarizabilities it is the TE$_{11}$ mode, in each case the lowest-order mode in the short section of circular waveguide forming the hole whose polarization matches that of the driving field. Akhiezer \cite{akh1}-\cite{akh2} has given expressions for the polarizabilities that are more uniformly valid over all values of $h/r_0$. By adapting his results to roughly track with the limiting forms (\ref{aE})-(\ref{aEl}) and (\ref{aM})-(\ref{aMl}), we can obtain the expressions
\begin{equation}
 \alpha_E^s \simeq \frac{2r_0^3}{3} \frac{1}{1 + 1.33 \tanh \left( \frac{2.405 h}{2r_0} \right)}
 \label{aEsu}
\end{equation}
\begin{equation}
 \alpha_E^a \simeq \frac{2r_0^3}{3} \frac{1}{1 + 1.33 \coth \left( \frac{2.405 h}{2r_0} \right)}
 \label{aEau}
\end{equation}
\begin{equation}
 \alpha_M^{s,tt} \simeq \frac{4r_0^3}{3} \frac{1}{1 + 1.817 \tanh \left( \frac{1.841 h}{2r_0} \right)}
 \label{aMsu}
\end{equation}
\begin{equation}
 \alpha_M^{a,tt} \simeq \frac{4r_0^3}{3} \frac{1}{1 + 1.817 \coth \left( \frac{1.841 h}{2r_0} \right)}
 \label{aMau}
\end{equation}
Similar but not identical formulas can be found in \cite{spor2}, but an assessment of the relative accuracy of the various formulas does not seem to have been carried out.

\subsection*{\ref{apa}.2 Apertures of other shapes}

Based on the foregoing formulas for a circular aperture, we might conjecture the following approximate ones for the polarizabilities of an aperture of more general cross section (but with enough symmetry that $\dyadic{\boldsymbol{\alpha}}_M^{(s,a)}$ is diagonal and isotropic) in a thick PEC screen:
\begin{equation}
 \alpha_{E,M}^s \simeq \frac{\alpha_{E,M}^0}{1 + \left( \frac{\alpha_{E,M}^0}{\alpha_{E,M}^{\infty}} - 1 \right) \tanh \left( \frac{k_{c(E,M)} h}{2} \right)}
 \label{aEMsu}
\end{equation}
\begin{equation}
 \alpha_{E,M}^a \simeq \frac{\alpha_{E,M}^0}{1 + \left( \frac{\alpha_{E,M}^0}{\alpha_{E,M}^{\infty}} - 1 \right) \coth \left( \frac{k_{c(E,M)} h}{2} \right)}
 \label{aEMau}
\end{equation}
where $\alpha_{E,M}^0$ are the zero-thickness polarizabilities, $\alpha_{E,M}^{\infty}$ are the polarizabilities in the limit as $h \rightarrow \infty$, and $k_{c(E,M)}$ are the cutoff wavenumbers of the lowest-order TM and TE modes respectively of a metallic waveguide whose cross section is that of the aperture shape. Clearly these formulas will be exact in either of the limits $h \rightarrow 0$ or $h \rightarrow \infty$. Similar formulas for apertures of square shape have been given in \cite{spor1}.

For a square aperture of side $l$ in a PEC plane of zero thickness, Fabrikant \cite{fab1}-\cite{fab2} has given the values
\begin{equation}
 \alpha_{E}^0 = \frac{l^3}{6\sqrt{2}} \qquad \alpha_{M}^0 = \frac{2 l^3}{9 \ln (1 + \sqrt{2})} \label{alphasquare}
\end{equation}
which are accurate to within a few percent. Park and Eom \cite{eom2,eom3} have given a numerical method for evaluating the polarizabilities of a rectangular aperture in a thick PEC screen, and give numerical results for the electric polarizabilities. McDonald \cite{mcd2} has computed results for transmission between two cavities due to an aperture of rectangular shape in the common thick wall, while in \cite{nik} are given formulas for the transmission coefficient resulting from a normally incident plane wave at such a rectangular aperture in a thick PEC plane. From these together with (\ref{aEMsu})-(\ref{aEMau}) we can extract approximate information about the values of $\alpha_{E,M}^{\infty}$ for a rectangular aperture, which at present will have to serve in the absence of more directly computed values of $\alpha_{E,M}^{s,a}$. As an example, for a square aperture we can estimate from the numerical results in the references cited above the values
\begin{equation}
 \alpha_{M}^{\infty} \simeq 0.3 \alpha_{M}^0 ; \qquad \alpha_{E}^{\infty} \simeq 0.5 \alpha_{E}^0
 \label{appsquare}
\end{equation}
but the error here might be substantially larger than the values for the circular aperture case, and more precise values will have to await more careful computations. However, by comparing (\ref{tilpi}), (\ref{tilchi}) and (\ref{pichi}) in the limit of $a \ll d$ with (\ref{square}), (\ref{aEMau}) and (\ref{alphasquare}), we can deduce that
\[
\frac{\alpha_M^{\infty}}{\alpha_M^0} = \frac{1}{\displaystyle 1 + \frac{\pi^3}{18 \ln (1 + \sqrt{2})}} \simeq 0.3385 
\]
which adds credence to at least the first estimate of (\ref{appsquare}). In a similar way, using (\ref{circle}) for an array of circular holes instead of (\ref{square}), we obtain the estimate
\[
 \frac{\alpha_M^{\infty}}{\alpha_M^0} = \frac{1}{\displaystyle 1 + \frac{3.2\, j'_{11}}{\pi}} \simeq 0.3478 
\]
which compares favorably to (\ref{aMinf0}).

\section*{Acknowledgment} 

The authors are grateful to Dr. Chris Holloway of the National Institute of Standards and Technology for a number of useful discussions about this paper.


\begin{thebibliography}{99}

\bibitem{ekel} E.~F. Kuester, E. Liu and N. J. Krull, ``Average transition conditions for electromagnetic fields at a metascreen of vanishing thickness,'' manuscript submitted to \emph{IEEE Trans. Ant. Prop.}.

\bibitem{hk3} C.~L. Holloway, E.~F. Kuester, J.~A. Gordon, J. O'Hara, J. Booth and D.~R. Smith, ``An overview of the theory and applications of metasurfaces: The two-dimensional equivalents of metamaterials'', \emph{IEEE Ant. Prop. Mag.}, vol. 54, no. 2, pp. 10-35, April 2012.

\bibitem{matsu} T. Matsumoto, M. Suzuki and C. Funatsu, ``Effect of a grid on the cavity of an electron tube,'' \emph{Electron. Commun. Japan}, vol. 48, no. 6, pp. 26-29, 1965.

\bibitem{otoshi} T. Y. Otoshi, ``A study of microwave leakage through perforated flat plates,'' \emph{IEEE Trans. Micr. Theory Tech.}, vol. 20, pp. 235-236, 1972.

\bibitem{chen} C.-C. Chen, ``Transmission of microwave through perforated flat plates of finite thickness,'' \emph{IEEE Trans. Micr. Theory Tech.}, vol. 21, pp. 1-6, 1973. % Note that in formulas (6) and (7), tan h should be tanh and cot h should be coth.

\bibitem{mcphed} R. C. McPhedran and D. Maystre, ``On the theory and solar application of inductive grids,'' \emph{Appl. Phys.}, vol. 14, pp. 1-20, 1977.

\bibitem{lee} S.-W. Lee, G. Zarrillo and C.~L. Law, ``Simple formulas for transmission through periodic metal grids or plates,'' \emph{IEEE Trans. Ant. Prop.}, vol. 30, pp. 904-909, 1982.

\bibitem{kaplun} V. A. Kaplun, I. T. Kravchenko and D. A. Tsaryuk, ``Diffraction of electromagnetic waves by a periodic screen with holes of arbitrary shape,'' \emph{Radiotekhnika}, vol. 37, no. 9, pp. 32-37, 1982 [in Russian; Engl. transl. in \emph{Telecomm. Radio Eng.}, vol. 36-37, no. 9, pp. 72-76, 1982]. % Numerical only (mode-matching).

\bibitem{abajo} F. J. Garc\'{i}a de Abajo, J. J. S\'{a}enz, I. Campillo and J. S. Dolado, ``Site and lattice resonances in metallic hole arrays,'' \emph{Opt. Express}, vol. 14, pp. 7-18, 2006. % No GSTC's but dipole interaction model.

\bibitem{martin} L. Mart\'{i}n-Moreno and F. J. Garc\'{i}a-Vidal, ``Minimal model for optical transmission through holey metal films,'' \emph{J. Phys. Cond. Mat.}, vol. 20, art. 304214, 2008.

\bibitem{edmunds} J. D. Edmunds, E. Hendry, A. P. Hibbins, J. R. Sambles and I. J. Youngs, ``Multi-modal transmission of microwaves through hole arrays,'' \emph{Opt. Exp.}, vol. 19, pp. 13793-13805, 2011. % Numerical and experimental data for thick hole array.

\bibitem{garcia} F. J. Garc\'{i}a-Vidal, L. Mart\'{i}n-Moreno, T. W. Ebbesen and L. Kuipers, ``Light passing through subwavelength apertures,'' \emph{Rev. Mod. Phys.}, vol. 82, pp. 729-787, 2010.

\bibitem{eom} H. H. Park and H. J. Eom, ``Electromagnetic scattering from multiple rectangular apertures in a thick conducting screen,'' \emph{IEEE Trans. Ant. Prop.}, vol. 47, pp. 1056-1060, 1999.

\bibitem{eom3} H. J. Eom, \emph{Wave Scattering Theory: A Series Approach Based on the Fourier Transformation}. Berlin: Springer-Verlag, 2001, chapter 6.

\bibitem{prelim} E.~F. Kuester and E. Liu, ``Average transition conditions for electromagnetic fields at a perfectly conducting metascreen,'' Paper TH-UB.1P.2, \emph{2015 IEEE AP-S Symposium on Antennas and Propagation and URSI CNC/USNC Joint Meeting}. 19-24 July 2015, Vancouver, BC Canada.

\bibitem{kmh} E.~F. Kuester, M.~A. Mohamed, M. Piket-May and C.~L. Holloway, ``Averaged transition conditions for electromagnetic fields at a metafilm,'' \emph{IEEE Trans. Ant. Prop.}, vol. 51, pp. 2641-2651, 2003.

\bibitem{senior}  T.~B.~A. Senior, and J.~L. Volakis, \emph{Approximate Boundary Conditions in Electromagnetics}. London: Institution of Electrical Engineers, 1995.

\bibitem{hkdrs}  C. L. Holloway, E. F. Kuester and A. Dienstfrey, ``A homogenization technique for obtaining generalized sheet transition conditions for an arbitrarily shaped coated wire grating,'' \emph{Radio Science}, vol. 49, pp. 813-850, 2014.

\bibitem{kont2} M.~I. Kontorovich, V.~Yu. Petrun'kin, N.~A. Esepkina and M.~I. Astrakhan, ``The coefficient of reflection of a plane electromagnetic wave from a plane wire mesh,'' \emph{Radiotekh. Elektron.}, vol. 7, pp. 239-249, 1962 [in Russian; Engl. transl. in \emph{Radio Eng. Electron. Phys.}, vol. 7, pp. 222-231, 1962].

\bibitem{kont3} M.~I. Kontorovich, ``Averaged boundary conditions at the surface of a grating with a square mesh,'' \emph{Radiotekh. Elektron.}, vol. 8, pp. 1506-1515, 1963 [in Russian; Engl. transl. in \emph{Radio Eng. Electron. Phys.}, vol. 8, pp. 1446-1454, 1963].

\bibitem{ast1} M.~I. Astrakhan, ``Averaged boundary conditions at the surface of a lattice with rectangular cells,'' \emph{Radiotekh. Elektron.}, vol. 9, pp. 1507-1508, 1964 [in Russian; Engl. transl. in \emph{Radio Eng. Electron. Phys.}, vol. 9, pp. 1239-1241, 1964].

\bibitem{ast2} M.~I. Astrakhan, ``Reflection and screening properties of plane wire grids,'' \emph{Radiotekhnika (Moscow)}, vol. 23, no. 1, pp. 23-30, 1968 [in Russian; Engl. transl. in \emph{Telecomm. Radio Eng.}, vol. 22-23, no. 1, pp. 76-83, 1968].

\bibitem{ast3} M.~I. Astrakhan, N.~M. Zolotukhina and G.~A. Sebyakina, ``Average boundary conditions for metal grids with conductors of noncircular cross section,'' \emph{Radiotekh. Elektron.}, vol. 20, pp. 2417-2420, 1975 [in Russian; Engl. transl. in \emph{Radio Eng. Electron. Phys.}, vol. 20, no. 11, pp. 154-156, 1975].

\bibitem{kont4} M.~I. Kontorovich, M.~I. Astrakhan, V.~P. Akimov and G.~A. Fersman, \emph{Elektrodinamika Setchatykh Struktur}. Moscow: Radio i Svyaz', 1987.

\bibitem{sivov} A. N. Sivov, A. D. Chuprin and A. D. Shatrov, ``Frequency independence effect of the phases of the reflection and transmission coefficients for short period gratings,'' \emph{Radiotekh. Elektron.}, vol. 39, pp. 1276-1278, 1994 [in Russian; Engl. transl. in \emph{J. Commun. Technol. Electron.}, vol. 39, no. 13, pp. 7-9, 1994].

\bibitem{chuprin} A. D. Chuprin, E. A. Parker, A. D. Shatrov, A. N. Sivov, V.~S. Solosin, A.~S. Zubov and R.~J. Langley, ``Phase characteristics of thick metal gratings,'' \emph{IEE Proc. pt. H}, vol. 145, pp. 411-415, 1998.

\bibitem{hkscreen} C. L. Holloway and E. F. Kuester, `Generalized sheet transition conditions for a metascreen---a fishnet metasurface,'' \emph{IEEE Trans. Ant. Prop.}, vol. 66, pp. 2414-2427, 2018.

\bibitem{medina} F. Medina, F. Mesa and R. Marqu\'{e}s, ``Extraordinary transmission through arrays of electrically small holes from a circuit theory perspective,'' \emph{IEEE Trans. Micr. Theory Tech.}, vol. 56, pp. 3108-3120, 2008.

% Single apertures in thick screen.

\bibitem{marc} N. Marcuvitz, \emph{Waveguide Handbook}. New York: McGraw-Hill, 1951, section 8.10.

\bibitem{smythe} W. R. Smythe, ``Flow over thick plate with circular hole,'' \emph{J. Appl. Phys.}, vol. 23, pp. 447-452, 1952.

\bibitem{akh1} A.~N. Akhiezer, ``On the inclusion of the effect of the thickness of the screen in certain diffraction problems,'' \emph{Zh. Tekh. Fiz.}, vol. 27, pp. 1294-1300, 1957 [in Russian; Engl. transl. in \emph{Sov. Phys. Tech. Phys.}, vol. 2, pp. 1190-1196, 1957].

\bibitem{akh2} A.~N. Akhiezer, ``On the coupling of rectangular waveguides by means of an aperture in the wide wall,'' \emph{Zh. Tekh. Fiz.}, vol. 30, pp. 851-854, 1960 [in Russian; Engl. transl. in \emph{Sov. Phys. Tech. Phys.}, vol. 5, pp. 802-805, 1961].

\bibitem{mcd2} N.~A. McDonald, ``Electromagnetic coupling through small apertures,'' Res. Rept. No. 45, Dept. Elec. Eng., Univ. of Toronto, 1971, chapter 6.

\bibitem{mcd} N.~A. McDonald, ``Electric and magnetic coupling through small apertures in shield walls of any thickness,'' \emph{IEEE Trans. Micr. Theory Tech.}, vol. 20, pp. 689-695, 1972.

\bibitem{spor1} F. Sporleder, \emph{Erweiterte Theorie der Lochkopplung}. Dr.-Ing. dissertation, Technische Universit\"{a}t Braunschweig, 1976, chapter V.

\bibitem{shap} E. A. Shapoval, ``Resonant frequencies of a cavity with a cylindrical hole in a wall of infinite thickness,'' \emph{Electron. Lett.}, vol. 14, pp. 445-447, 1978.

\bibitem{spor2} F. Sporleder and H.-G.Unger, \emph{Waveguide Tapers, Transitions and Couplers}. Stevenage, UK: Peter Peregrinus, 1979, pp. 51-52.

\bibitem{gluck} R.~L. Gluckstern and J.~A. Diamond, ``Penetration of fields through a circular hole in a wall of finite thickness,'' \emph{IEEE Trans. Micr. Theory Tech.}, vol. 39, pp. 274-279, 1991.

\bibitem{eom2} H. H. Park and H. J. Eom, ``Electrostatic potential distribution through a rectangular aperture in a thick conducting plane,'' \emph{IEEE Trans. Micr. Theory Tech.}, vol. 44, pp. 1745-1747, 1996.

\bibitem{nik} A.~Yu. Nikitin, D.~Zueco, F.~J. Garc\'{i}a-Vidal, and L. Mart\'{i}n-Moreno, ``Electromagnetic wave transmission through a small hole in a perfect electric conductor of finite thickness,'' \emph{Phys. Rev. B}, vol. 78, art. 165429, 2008. %Numerical solution for single holes, fitted analytical formulas.

\bibitem{fab1} V.~I. Fabrikant, ``Magnetic polarizability of small apertures: analytical approach,'' \emph{J. Phys. A:  Math. Gen.}, vol. 20, pp. 323-338, 1987.

\bibitem{fab2} V.~I. Fabrikant, ``Electrical polarizability of small apertures: analytical approach,'' \emph{Int. J. Electron.}, vol. 62, pp. 533-545, 1987.

\end{thebibliography}
\end{document}